\documentclass[aip,preprint,longbibliography]{revtex4-1}
\usepackage{graphicx}
\usepackage{comment}
\usepackage{xcolor}
\usepackage[colorlinks=true,allcolors=blue]{hyperref}

\draft 

\begin{document}


\title{Ion beam sputtering of silicon: Energy distributions of sputtered and scattered ions} 



\author{Dmitry Kalanov}
\email[Email: ]{dmitry.kalanov@iom-leipzig.de}
\affiliation{Leibniz Institute of Surface Engineering (IOM), Permoserstraße 15, 04318 Leipzig, Germany}
\author{Andr\'e Anders}
\affiliation{Leibniz Institute of Surface Engineering (IOM), Permoserstraße 15, 04318 Leipzig, Germany}
\affiliation{Felix Bloch Institute of Solid State Physics, Leipzig University, Linnéstraße 5, Leipzig, Germany}

\author{Carsten Bundesmann}
\affiliation{Leibniz Institute of Surface Engineering (IOM), Permoserstraße 15, 04318 Leipzig, Germany}

\date{\today}

\begin{abstract}
The properties of sputtered and scattered ions are studied for ion beam sputtering of Si by bombardment with noble gas ions. The energy distributions in dependence on ion beam parameters (ion energy: 0.5 - 1 keV; ion species: Ne, Ar, Xe) and geometrical parameters (ion incidence angle, polar emission angle, scattering angle) are measured by means of energy-selective mass spectrometry. The presence of anisotropic effects due to direct sputtering and scattering is discussed and correlated with process parameters. The experimental results are compared to calculations based on a simple elastic binary collision model and to simulations using the Monte-Carlo code SDTrimSP. The influence of the contribution of implanted primary ions on energy distributions of sputtered and scattered particles is studied in simulations. It is found that a 10\% variation of the target composition leads to detectable but small differences in the energy distributions of scattered ions. Comparison with previously reported data for other ion/target configurations confirms the presence of similar trends and anisotropic effects: the number of high-energy sputtered ions increases with increasing energy of incident ions and decreasing scattering angle. The effect of the ion/target mass ratio is additionally investigated. Small differences are observed with the change of the primary ion species: the closer the mass ratio to unity, the higher the average energy of sputtered ions. The presence of peaks, assigned to different mechanisms of direct scattering, strongly depends on the ion/target mass ratio.
\end{abstract}


\maketitle 

\section{Introduction}
Ion beam sputter deposition (IBSD) is one of the well-known physical vapor deposition (PVD) techniques. In comparison to other PVD techniques, such as magnetron sputtering or evaporation, IBSD offers the opportunity to tailor the properties of film-forming particles, and, therefore, film properties by varying the parameters of ion beam and sputtering geometry over a wide range \cite{Behrisch2007}. In the IBSD system, the generation of energetic ions (ion beam source), formation of secondary particle fluxes (sputtered and scattered ones at the target), and the film growth area are locally separated. Change of ion beam parameters and sputtering geometry results in different angular and energy distributions of sputtered target and scattered primary particles. The working pressure is typically less than $10^{-2}$ Pa, so the mutual interaction within the secondary particle fluxes is much weaker than in magnetron sputtering or evaporation.

Although IBSD has been used for decades, its full capabilities have not been studied systematically until recently. Typically, a beam of primary ions with energies up to 2 keV is used. Several groups performed angularly resolved measurements of energy/velocity distributions of sputtered atoms \cite{Goehlich1996,Franke1997,Goehlich1997,Goehlich1999,Goehlich2000,Goehlich2000a,Goehlich2001,Stepanova2001,Stepanova2002,Stepanova2002a,Stepanova2004}  in the mentioned energy range of primary ions. They reported deviations from the well-known Thompson formula \cite{Thompson1968,Sigmund1969}, which were related to ion energy, ion incidence angle and ion species \cite{Stepanova2001,Stepanova2002,Stepanova2002a}, or to the target topography \cite{Stepanova2004}. At the same time, the influence of scattered primary ions was usually neglected. In recent works, it was shown that these ions can have high energy and a large impact on thin film properties \cite{Lautenschlager2017,Mateev2018}. Previously, systematic investigations of ion beam sputtering were made for Ag, Ge, and Ti targets, bombarded by Ar, Xe or O$_2$ ions. The trends and correlations between process parameters, particle and film properties for those IBSD configurations were discussed in a recent tutorial \cite{Bundesmann2018}.

A quite appealing material for IBSD studies is Si, because of its importance in photovoltaics and microelectronics. Since systematic trends and differences in properties of thin films, produced with IBSD, were attributed to secondary particle properties \cite{Bundesmann2018}, the measurements of their energy distributions are essential steps towards the deeper understanding of the process. There are some reports on the angular distributions \cite{Okutani1980,Chernysh2004} and energy distributions \cite{Wittmaack1984} of Si sputtered by 1-3 keV Ar ions. However, the energy distributions of sputtered Si ions were measured only up to 80 eV. The present study focuses on energy distributions of sputtered and scattered ions in a significantly wider energy range, where needed up to 1000 eV, measured during bombardment of a Si target with primary ions of Ne, Ar and Xe. Measurements using energy-selective mass spectrometry are supplemented by simulations using the Monte-Carlo code SDTrimSP \cite{Mutzke2011} and calculations based on a simple elastic binary collision model. The correlations between parameters and properties, similarities and differences from other ion/target configurations are discussed.

\section{Experiment}
\begin{figure}[t!]
\includegraphics[width = 0.75\linewidth]{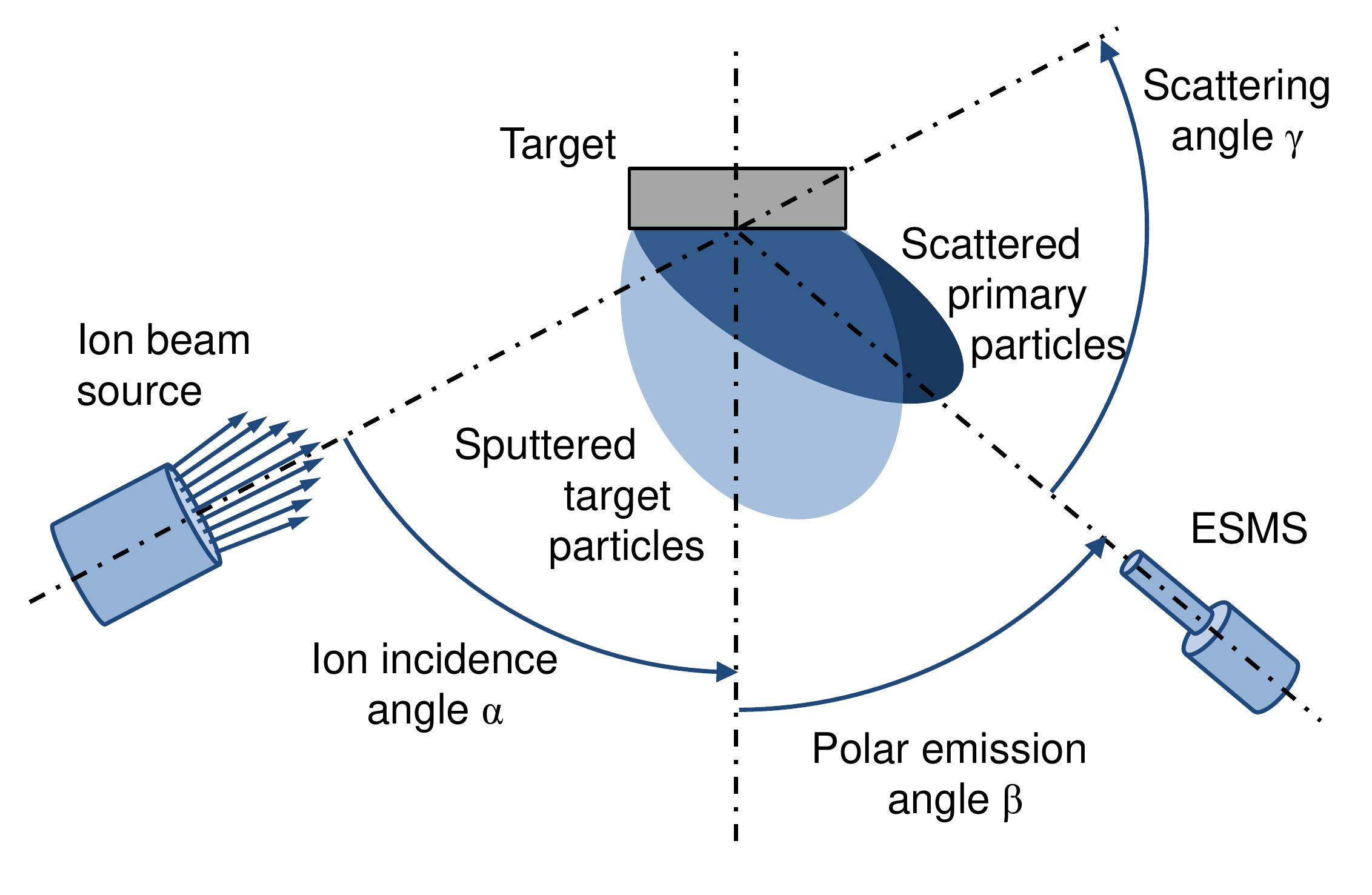}
\centering
\caption{Scheme of the experimental setup.} 
\label{fig:setup}
\end{figure}

Figure \ref{fig:setup} shows a schematic sketch of the ion beam sputter setup inside the vacuum chamber. The setup consists of a broad-beam ion source, a target holder and an energy-selective mass spectrometer (ESMS).

The ion beam source is based on a radio-frequency (RF) discharge with a three-grid multi-aperture extraction system with an open diameter of 16 mm \cite{Zeuner2001}. Process gases were Ne, Ar, and Xe with a mass flow rate of 20 sccm, 3.5 sccm and 1.3 sccm respectively. The base pressure in the vacuum chamber was $1.5\times 10^{-4}$ Pa, and the typical process pressure was about $5\times 10^{-3}$ Pa. The RF-power was set to 105 W for Ne, 70 W for Ar, and 35 W for Xe, producing a total ion current of about 6 mA in all cases. The target was polycrystalline Si (diameter 100 mm, thickness 3 mm, purity 99.999\%). 

Energy distributions of sputtered and scattered ions were measured using the energy-selective mass spectrometer (ESMS) Hiden EQP. The differentially pumped ESMS was operated in the energy range from 0 eV up to 1000 eV. The ESMS was mounted at the chamber wall, and the orifice diameter was 0.3 mm.

The target and the ion source are mounted on rotary tables with a common rotation axis. The setup provided a possibility to vary the ion incidence angle $\alpha$ and the polar emission angle $\beta$. The energy distributions of sputtered Si and scattered Ne, Ar and Xe singly charged ions were measured at emission angles in steps of $20^\circ$ for primary ion energies of 0.5 keV and 1.0 keV, and incidence angles of $30^\circ$ and $60^\circ$. The yields of multiply charged ions are orders of magnitude lower, and they were not considered.

\section{Modeling and calculations}
Energy distributions were also simulated using the Monte Carlo code SDTrimSP \cite{Mutzke2011}. This code is widely used for static and dynamic simulations of the sputtering process in various materials. It is based on the binary collision approximation when each collision between atoms is described as an elastic collision with an interaction potential. The energy transfer to electrons is considered as a separate inelastic energy loss.

All simulations were performed with $10^8$ primary ions to achieve sufficient statistics. Ion-solid interactions were described using the Kr-C potential \cite{Wilson1977}, and the inelastic energy loss model was equipartition of Lindhard-Scharff \cite{Lindhard1961} and Oen-Robinson models \cite{Oen1976}. The composition of the target for static simulations was assumed to be 90\% Si and 10\% implanted particles. These numbers are estimates of the average fraction of implanted primary ions in the near-surface layers of the target. The estimates were obtained by dynamic SDTrimSP simulations. The influence of scattering at implanted ions was previously observed for different target materials \cite{Feder2013,Bundesmann2015,Feder2014,Lautenschlager2016}. Please note that the code does not account for the formation of secondary ions. We use the resulting simulated distributions as being comprised of the sum of neutral atoms and ions. Later in the text, we will use the term \emph{particle} to describe either a neutral atom or an ion where we cannot distinguish between them.

The sputtering process in the energy range around 1 keV is commonly described by redistribution of momentum and energy through linear collision cascades. At the same time, target atoms can be sputtered after a single collision, which is designated as \emph{directly sputtered}. Similarly, the incoming ions can be \emph{directly scattered} at particles of the surface or near-surface region. Those events can be identified as features in experimental and simulated energy distributions \cite{Feder2013,Bundesmann2015,Feder2014,Lautenschlager2016}. 

Direct sputtering and scattering can be described in a simplified manner by the conservation of momentum and energy in a binary elastic collision. With the incoming primary ion of mass $M_{\text{ion}}$ and initial energy $E_{\text{ion}}$, and a target atom with mass $M_{\text{tar}}$ at rest ($E_{\text{tar}}=0$), the energy of sputtered target particle $E_{\text{spu}}$ and scattered primary particle $E_{\text{sca}}$ are calculated as
\begin{eqnarray}
\label{eq:direct_spu}
&&E_{\text{spu}} = E_{\text{ion}}\frac{4 M_{\text{ion}} M_{\text{tar}}}{\left(M_{\text{ion}} + M_{\text{tar}}\right)^2} \cos^2(\gamma),\\
\label{eq:direct_sca}
&&E_{\text{sca}} = E_{\text{ion}}\left(\frac{M_{\text{ion}}\cos(\gamma) + \sqrt{M^2_{\text{tar}} - M^2_{\text{ion}}\sin^2(\gamma)}}{M_{\text{ion}} + M_{\text{tar}}}\right)^2,
\end{eqnarray}
where $\gamma = 180^\circ - \alpha - \beta$ denotes the scattering angle. If $M_{\text{ion}} > M_{\text{tar}}$, the scattering angles are restricted by $\gamma_{\text{max}} = \arcsin(M_{\text{tar}}/M_{\text{ion}}) < 90^\circ$. In case of Ar-Si interaction ($M_{\text{ion}} = 39.95$ u, $M_{\text{tar}} = 28.09$ u), the limit is $\gamma_{\text{max}}\approx 44.4^\circ$. If $M_{\text{ion}} < M_{\text{tar}}$,  angles $\gamma \leq 180^\circ$ are allowed for scattered particles, and $\gamma \leq 90^\circ$ for sputtered particles.

\begin{table*}[t!]
\caption{\label{tab:direct_event_calc} Calculated energies for the direct sputtering/scattering events at incidence angles $\alpha = 60^\circ$ and  $40^\circ$ and different emission angles. Scattering angle $\gamma = 180^\circ - \alpha - \beta$ is used in the table.}
\begin{ruledtabular}
\begin{tabular}{cccccccccc}
& & Ne-Si & & Ar-Si & & Xe-Si & & Ar-Ar; Ne-Ne; Xe-Xe \\ \cline{3-9}
$E_{\text{ion}}$ [keV] & $\gamma$ [$^\circ$] & $E_{\text{sca}}$ [eV] & $E_{\text{spu}}$ [eV] & $E_{\text{sca}}$ & $E_{\text{spu}}$ & $E_{\text{sca}}$ & $E_{\text{spu}}$ & $E_{\text{sca}} = E_{\text{spu}}$ \\
\hline
0.5	& 100 & 57 & - & - & - & - & - & - \\
&	80 & 117 & 15 & - &	15 & - & 9 & 15 \\
&	60 & 221 & 122 & - & 121 & - & 73 & 125 \\
&	40 & 350 & 286 & 188 & 284 & - & 170 & 293 \\
1.0 & 100 & 115 & - & - & - & - & - & - \\
& 	80 & 234 & 29 & - & 29 & - & 30 & 30 \\
&	60 & 442 & 243 & - & 242 & - & 250 & 250 \\
&	40 & 700 & 571 & 377 & 569 & - & 587 & 587 \\
	
\end{tabular}
\end{ruledtabular}
\end{table*}

Eqs. \ref{eq:direct_spu} and \ref{eq:direct_sca} describe the ideal case. In reality, incident ions face additional energy losses. First, there is the inelastic energy loss to electrons of the target, the so-called "electronic stopping". It depends on the energy of incident ion and geometry. Assuming that direct sputtering happens in the 2-3 topmost atomic layers of the target, calculations by SRIM \cite{SRIM2013} yield values around 7 eV for Ne-Si, 11 eV for Ar-Si and 8 eV for Xe-Si interaction.
Second, sputtered and scattered particles have to overcome the surface potential, described by the surface binding energy $E_{\text{sb}}$. For Si $E_{\text{sb}}$ is determined to be 4.66 eV \cite{Mutzke2011}, and for inert gas atoms the surface binding energy is close to zero.

Table \ref{tab:direct_event_calc} contains calculated energies of direct sputtering and scattering events between an incident primary ion and a Si target atom or an implanted primary ion using Eqs. (\ref{eq:direct_spu}) and (\ref{eq:direct_sca}). Considering Ar-Ar, Ne-Ne and Xe-Xe interaction, Equations (\ref{eq:direct_sca}) and (\ref{eq:direct_spu}) become identical. That means that sputtering and scattering events cannot be distinguished in the experimental energy distribution in this case.

\section{Results and discussion}
\subsection{Energy distributions of sputtered Si ions}
\begin{figure*}[t!]
\includegraphics[width = 1\linewidth]{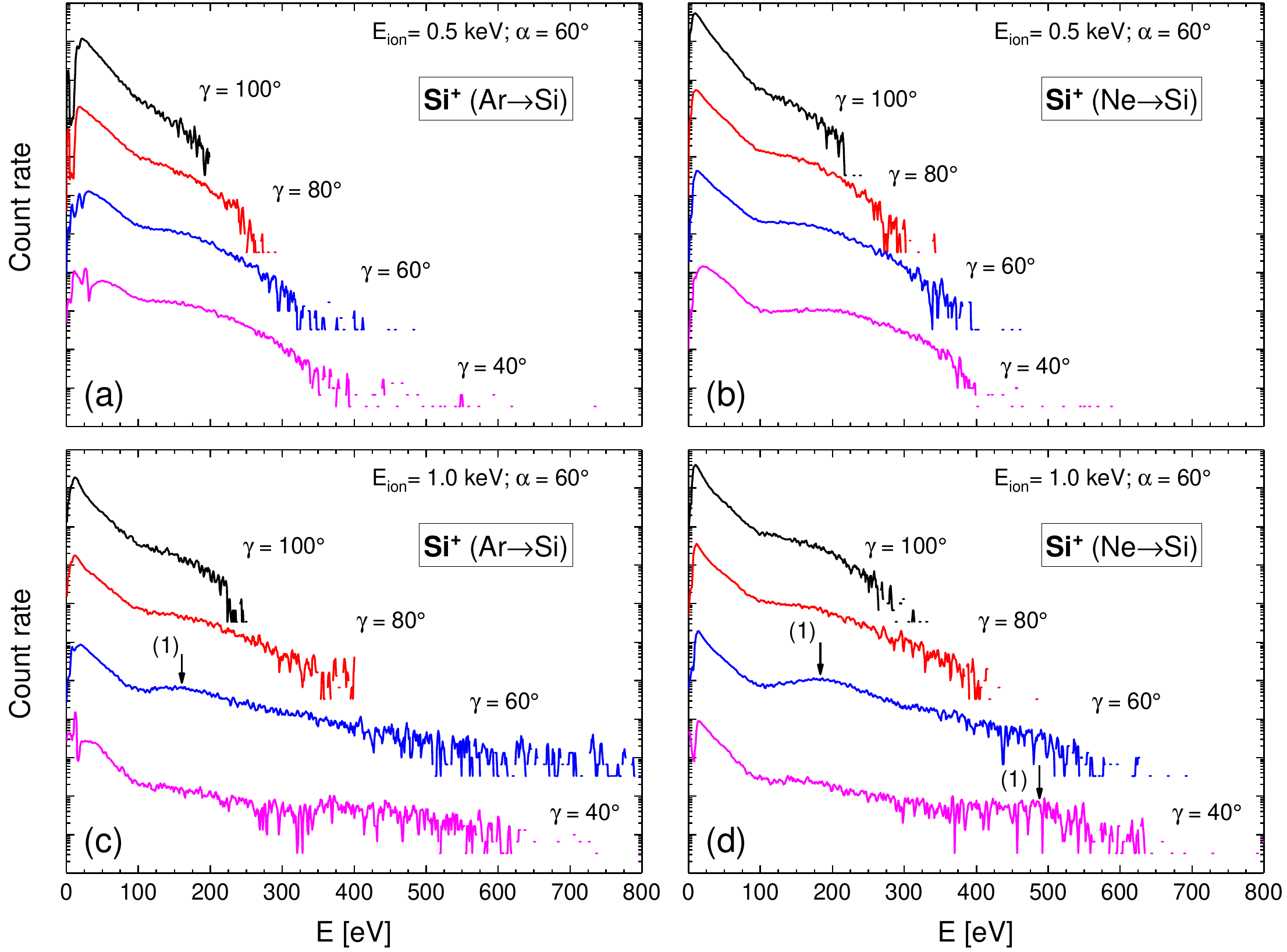}
\centering
\caption{Measured energy distributions of Si ions sputtered by bombardment with Ar (a,c) and Ne (b,d) ions in dependence on the scattering angle $\gamma$. Panels show the influence of the ion energy ($E_{\text{ion}}= 0.5$ keV, 1.0 keV) at a fixed ion incidence angle $\alpha = 60^\circ$.  The arrows indicate peaks assigned to direct sputtering (1).} 
\label{fig:Spu_Si_exp}
\end{figure*}

Figure \ref{fig:Spu_Si_exp} shows measured energy distributions of sputtered ions for the bombardment of a Si target with Ar and Ne ions. All distributions have a low-energy peak followed by a decay proportional to $E^{-n}$ up to 100 eV. 

At higher energies, a high energy tail evolves. With decreasing scattering angle, the tail becomes more prominent. The presence of the tail is explained by anisotropy effects \cite{Goehlich1996,Franke1997,Goehlich1997,Goehlich1999,Goehlich2000,Goehlich2000a,Goehlich2001,Stepanova2001,Stepanova2002,Stepanova2002a,Stepanova2004}.  At scattering angles $\gamma = 60^\circ$ and  $40^\circ$, broad structures at high energies can be seen (marked by arrows). In previous studies with other target materials \cite{Feder2013,Bundesmann2015,Feder2014,Lautenschlager2016}, this structure was associated with direct sputtering since the position of its maximum agreed well with the energy of a direct event (Eq. (\ref{eq:direct_spu})). The label near the arrow in Figure refers to the cited equation.

Figs. \ref{fig:Spu_Si_exp}(a,c) and \ref{fig:Spu_Si_exp}(b,d) show distributions for various incident ion energies. The overall distribution shape stays almost the same if the same scattering angle is considered. Broad features and the high-energy tail are linearly shifted to higher energies, the distribution is "stretched" when incident ions have higher energy. 

With the change of the process gas from Ar to Ne (Figs. \ref{fig:Spu_Si_exp}(a,b) and \ref{fig:Spu_Si_exp}(c,d)), the shapes of the distributions change only slightly.

\begin{figure*}[t!]
\includegraphics[width = 1\linewidth]{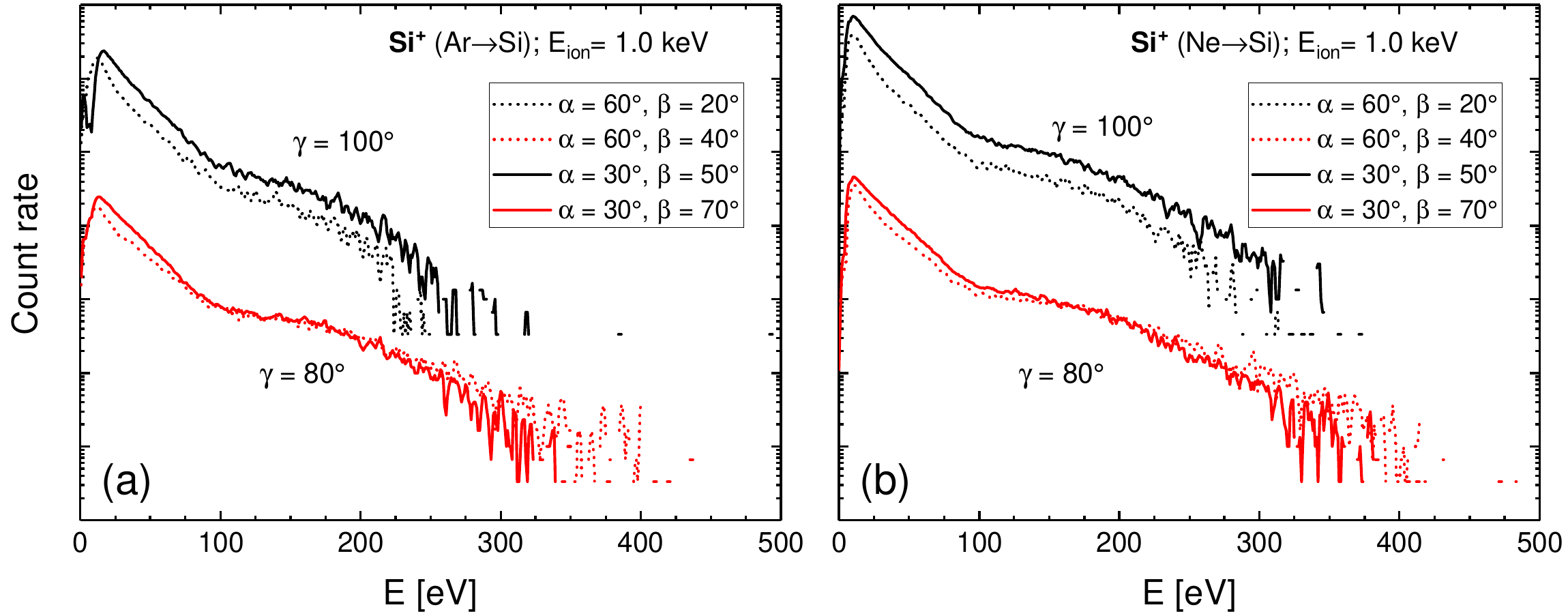}
\centering
\caption{Measured energy distributions of Si ions sputtered by bombardment with Ar ions (a) and Ne ions (b) in dependence on the scattering angle $\gamma$. Panels show the influence of the incidence angle ($\alpha = 60^\circ$, $30^\circ$) at a fixed ion incidence energy $E_{\text{ion}}= 1.0$ keV.} 
\label{fig:Si_spu_angle}
\end{figure*}

The distributions for various incidence angles in Figure \ref{fig:Si_spu_angle} show that the shape of the distributions changes with the scattering angle $\gamma$, and only slightly depends on specific combinations of incident angle $\alpha$ and emission angle $\beta$.

\subsection{Energy distributions of scattered ions}
\begin{figure*}[t!]
\includegraphics[width = 1\linewidth]{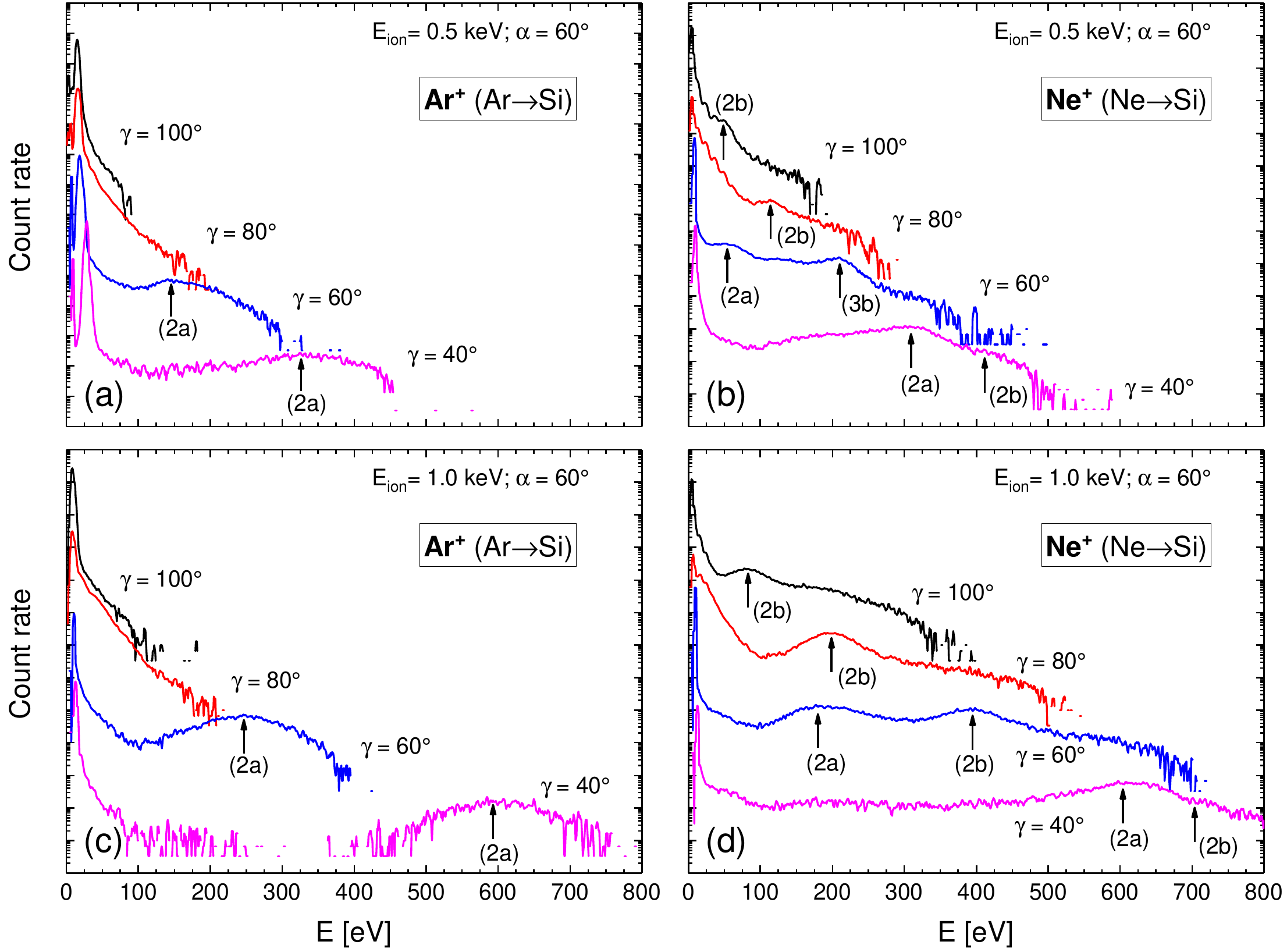}
\centering
\caption{Measured energy distributions of Ar (a,c) and Ne (b,d) ions scattered and sputtered from a Si target in dependence on the scattering angle $\gamma$. Panels show the influence of the ion energy ($E_{\text{ion}}= 0.5$ keV, 1.0 keV) at a fixed ion incidence angle ($\alpha = 60^\circ$). The arrows indicate peaks assigned to direct scattering on implanted primary ions (2a), Si atoms (2b).} 
\label{fig:Sca_Ar_exp}
\end{figure*}

Figure \ref{fig:Sca_Ar_exp} shows measured energy distributions of scattered Ar and Ne ions. The low-energy maximum is followed by an abrupt signal drop. The high-energy tail also extends to higher energies with decreasing scattering angles. At scattering angles $\gamma  = 60^\circ$ and $40^\circ$ a broad structure with a local maximum appears. This structure (marked by arrows and the number (2a)) is assigned to direct scattering at implanted ions because its energy agrees well with the calculated energy according to the Table \ref{tab:direct_event_calc}. In the case of Ne scattering, a second peak shows up at higher energies. This peak is related to direct Ne-Si scattering (Eq. (\ref{eq:direct_sca})). Direct Ar-Si scattering can only occur for $\gamma < 44.4^\circ$ due to the higher atomic mass of Ar, and is indistinguishable from the noise. Implanted primary particles can be directly sputtered as well, but the energy position will coincide with Ar-Ar/Ne-Ne scattering and can therefore not be distinguished from these. Also, lighter Ne ions are scattered more efficiently than heavier Ar ions, therefore showing a more prominent high-energy tail in all cases.

The dependency of energy distributions on incident ion energy (Figs. \ref{fig:Sca_Ar_exp}(a,c) and \ref{fig:Sca_Ar_exp}(b,d)) is similar to that for sputtered atoms. At higher incident ion energies, the direct scattering features and high-energy tail are shifted to higher energies. The peak positions of the direct event features are in good agreement with the simple calculation (Table \ref{tab:direct_event_calc}).

\begin{figure*}[t!]
\includegraphics[width = 1\linewidth]{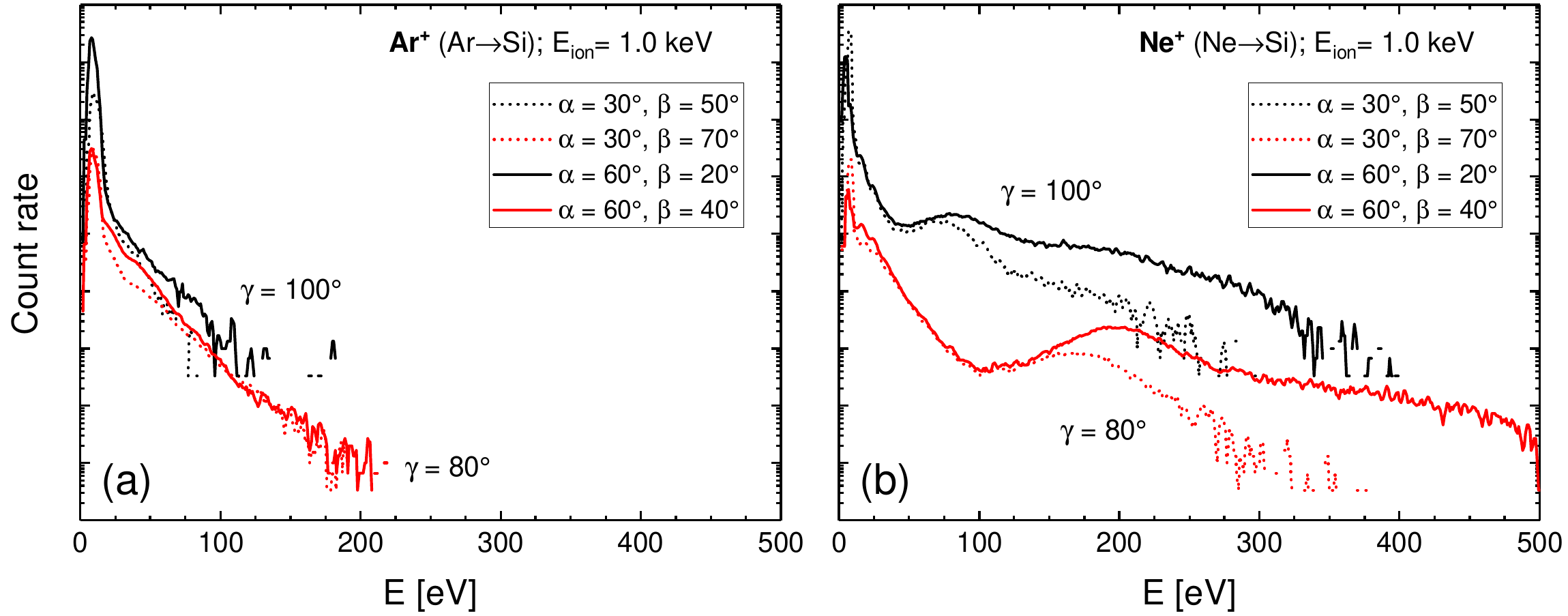}
\centering
\caption{Measured energy distributions of Ar ions (a) and Ne ions (b) scattered and sputtered from a Si target in dependence on the scattering angle $\gamma$. Panels show the influence of the incidence angle ($\alpha = 60^\circ$, $30^\circ$) at a fixed ion incidence energy ($E_{\text{ion}}= 1.0$ keV).} 
\label{fig:sca_angle}
\end{figure*}

Distributions of scattered Ar ions are also almost independent of specific combinations of incidence and emission angles when curves for the same scattering angle are compared (Figure \ref{fig:sca_angle}a). However, light Ne ions are directly scattered with higher energies at higher incidence angles (which means lower emission angles) than the heavier Ar ions, as can be seen from  Figure \ref{fig:sca_angle}b. This feature correlates with an increasing role of anisotropy effects for lower scattering angles and lower mass ratios, as was reported in Refs. \cite{Roosendaal1980,Lautenschlager2016}.

\subsection{Simulated distributions of secondary particles}

\begin{figure*}[t!]
\includegraphics[width = 1\linewidth]{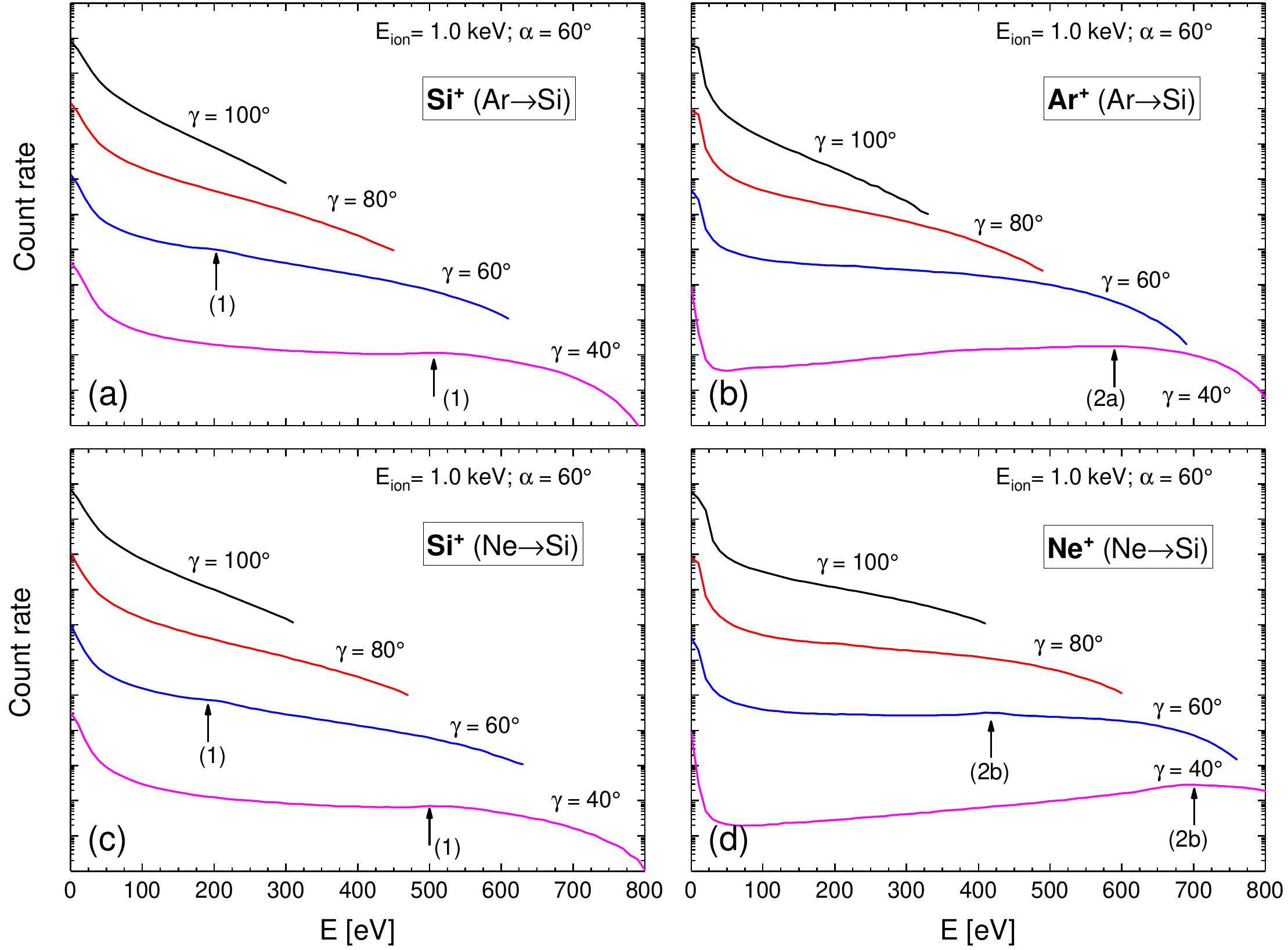}
\centering
\caption{Simulated energy distributions of secondary particles in dependence on the scattering angle $\gamma$ at a fixed ion incidence angle $\alpha = 60^\circ$ and incident ion energy $E_{\text{ion}}= 1.0$ keV, sputtered and scattered from Si$_{0.9}$Ar$_{0.1}$ (a,b) and Si$_{0.9}$Ne$_{0.1}$ (c,d) target. Panels (a,c) show sputtered Si particles, panels (b,d) show scattered Ar (b) and Ne (d) particles. The arrows indicate peaks assigned to direct sputtering (1), direct Ar-Ar scattering (2a), and direct Ne-Si scattering (2b).}
\label{fig:Spu_Si_calc}
\end{figure*}

The simulated energy distributions in Figure \ref{fig:Spu_Si_calc} qualitatively reproduce the systematic trends of the measured data, namely peaks of direct scattering and sputtering, and decreasing steepness of the high-energy tail at smaller scattering angles. The positions of resolved peaks agree well with experimental findings and calculations using the simple model (Eqs. \ref{eq:direct_spu}, \ref{eq:direct_sca}).

\begin{figure*}[t]
\includegraphics[width = 1\linewidth]{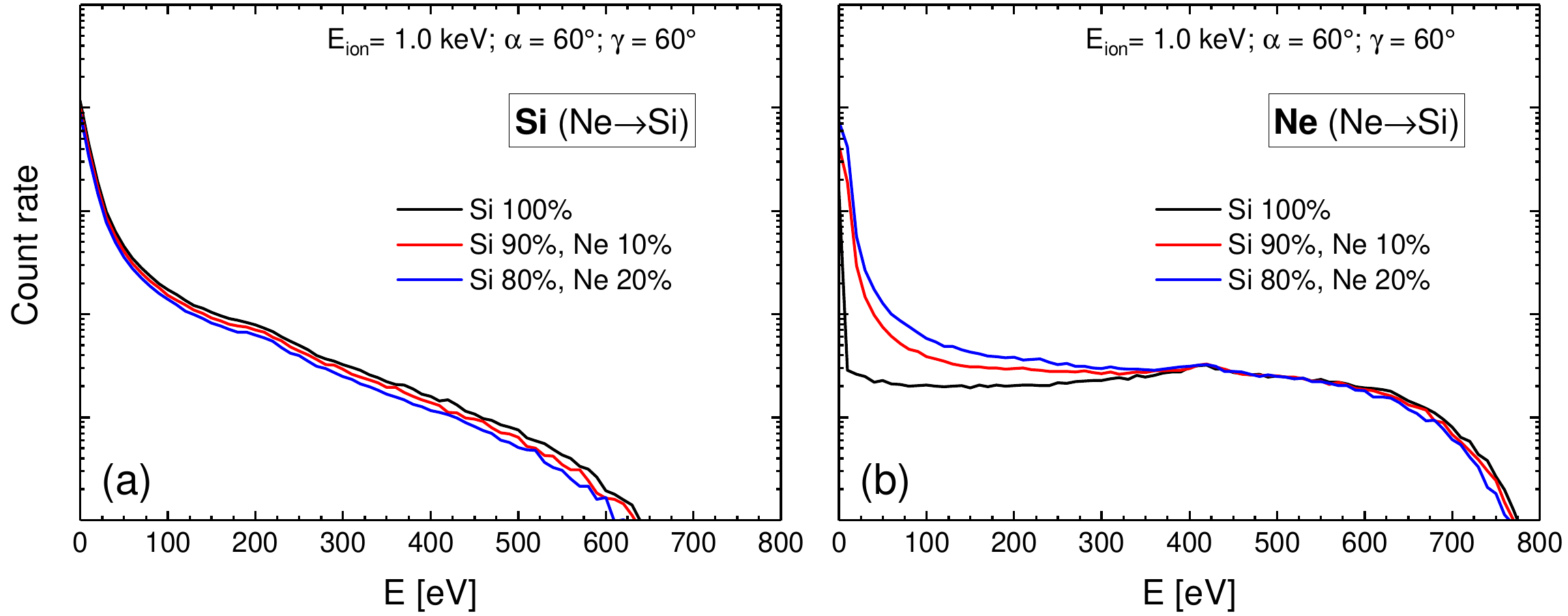}
\centering
\caption{Simulated energy distributions of sputtered Si (a) and scattered Ne (b) particles in dependence on the target composition at a fixed ion incidence angle $\alpha = 60^\circ$, scattering angle $\gamma = 60^\circ$ and incident ion energy $E_{\text{ion}}= 1.0$ keV.}
\label{fig:SiNe_varcomp}
\end{figure*}

Previously it was specified that the target composition was assumed to contain 90\% Si and 10\% implanted primary ions, based on the estimates of dynamic simulations. It is interesting to vary the percentage of implanted ions in order to study its effect on the energy distributions of sputtered and scattered particles. Figure \ref{fig:SiNe_varcomp} shows the case of Ne-Si bombardment at a fixed scattering angle $\gamma = 60^\circ$ for three target compositions: Si$_{1.0}$, Si$_{0.9}$Ne$_{0.1}$ and Si$_{0.8}$Ne$_{0.2}$. Sputtered particles show a slight dependence on the gas contribution. Scattered primary particles show a small increase in numbers in the low-energy part of the distribution with an increase of the gas contribution to the target. This increase is assigned to sputtering of previously implanted particles. The high energy tail stays the same, preserving the features of direct scattering. 

\subsection{Discussion}

It is interesting to consider systematic correlations between the present data and the results for Ag \cite{Feder2013,Bundesmann2015}, Ge \cite{Feder2014} and Ti \cite{Lautenschlager2016}, sputtered by Ar and Xe ions.

In all experiments, the average energy of the sputtered ions increases with decreasing scattering angle and increasing incident ion energy. The change of the primary ion species affects the distributions only slightly. The more important factor is the mass ratio between primary ion and target atom. The closer the ratio to unity, the higher is the maximum energy transferrable in a direct sputtering event (Eq. (\ref{eq:direct_spu}) with $\gamma = 0^\circ$). 

In distributions of scattered ions, the broad peaks, assigned to direct scattering at implanted primary ions, were observed for all primary ion/target combinations. Positions of their maxima are independent of the combination, since particles of the same mass interact, but the peaks are higher for lighter atoms. In Ar-Ag, Ar-Ge and Ne-Si configurations, energy distributions also have narrow peaks, assigned to direct scattering on target atoms. In several cases, when the mass ratio is close to unity, the structures assigned to different scattering processes are mixed \cite{Lautenschlager2016}.

\begin{figure*}[t!]
\includegraphics[width = 1\linewidth]{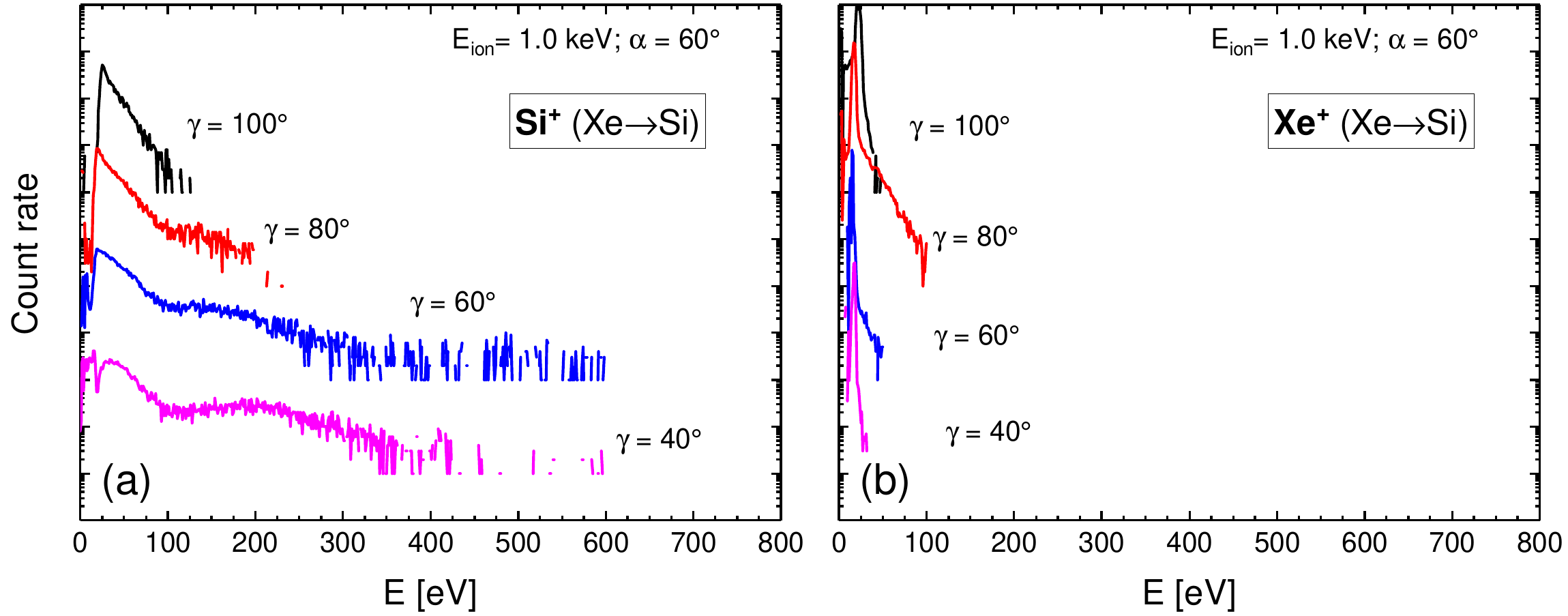}
\centering
\caption{Measured energy distributions of sputtered Si (a) and scattered Xe (b) ions in dependence on the scattering angle $\gamma$ at a fixed ion incidence angle $\alpha = 60^\circ$ and incident ion energy $E_{\text{ion}}= 1.0$ keV.} 
\label{fig:XeSi}
\end{figure*}

Those trends are additionally verified by measurements for Si bombarded by Xe ions (Figure \ref{fig:XeSi}). In contrast to Ne-Si and Ar-Si cases, the ion/target mass ratio is much higher than unity, the average energy of sputtered ions is smaller, and the scattering process is less efficient.

\section{Summary}

In the present work, the energy distributions of sputtered and scattered ions in the ion beam sputtering process of Si were studied experimentally by energy-selective mass spectrometry, and by simulations using the SDTrimSP code in dependence on ion beam parameters and sputtering geometry.

The distributions of sputtered Si ions show similar dependencies as in previous IBSD configurations with other target materials: the average energy of the sputtered ions increases with increasing ion energy and decreasing scattering angle. Also, small differences are observed with the change of the primary species: the closer the mass ratio to unity, the higher the average energy of sputtered ions. 

The distributions of scattered Ar ions show the importance of direct scattering on previously implanted primary ions and no signs of scattering at Si. This observation is supported by the fact that the Ar/Si mass ratio is higher than unity, therefore direct scattering happens only at oblique angles. At the same time, in case of the Ne-Si configuration, the distributions reveal the presence of both direct scattering channels. The average ion energy in both Ar and Ne energy distributions also increases with ion energy and decreasing scattering angle. Finally, lighter primary ions tend to scatter more efficiently and with higher average energy, as can be seen from the comparison of energy distributions of Ne$^+$, Ar$^+$ and Xe$^+$ ions. 

A comparison of experimental data with simulated energy distributions shows qualitative agreement and the same trends, and the presence of direct scattering and sputtering. Additional simulations show that a 10\% variation of the contribution of the target composition does not lead to significant differences in the energy distributions of sputtered and scattered particles.

The results presented in this paper complement systematic studies of IBSD with a very important material. In principle, the discussed trends and effects may also be applicable to other materials. Due to great technological relevance, future measurements in the presence of oxygen will complement the current and recent studies of SiO$_2$ films made with IBSD \cite{Mateev2018}.



%
%

%

\begin{acknowledgments}
The authors thank T. Amelal, R. Feder and F. Scholze for fruitful discussions. The software code SDTrimSP is kindly provided by the Max Planck Institute for Plasma Physics (IPP), Germany.
\end{acknowledgments}

\bibliography{D:/Science/Articles_temp/library}

\begin{thebibliography}{31}%
\makeatletter
\providecommand \@ifxundefined [1]{%
 \@ifx{#1\undefined}
}%
\providecommand \@ifnum [1]{%
 \ifnum #1\expandafter \@firstoftwo
 \else \expandafter \@secondoftwo
 \fi
}%
\providecommand \@ifx [1]{%
 \ifx #1\expandafter \@firstoftwo
 \else \expandafter \@secondoftwo
 \fi
}%
\providecommand \natexlab [1]{#1}%
\providecommand \enquote  [1]{``#1''}%
\providecommand \bibnamefont  [1]{#1}%
\providecommand \bibfnamefont [1]{#1}%
\providecommand \citenamefont [1]{#1}%
\providecommand \href@noop [0]{\@secondoftwo}%
\providecommand \href [0]{\begingroup \@sanitize@url \@href}%
\providecommand \@href[1]{\@@startlink{#1}\@@href}%
\providecommand \@@href[1]{\endgroup#1\@@endlink}%
\providecommand \@sanitize@url [0]{\catcode `\\12\catcode `\$12\catcode
  `\&12\catcode `\#12\catcode `\^12\catcode `\_12\catcode `\%12\relax}%
\providecommand \@@startlink[1]{}%
\providecommand \@@endlink[0]{}%
\providecommand \url  [0]{\begingroup\@sanitize@url \@url }%
\providecommand \@url [1]{\endgroup\@href {#1}{\urlprefix }}%
\providecommand \urlprefix  [0]{URL }%
\providecommand \Eprint [0]{\href }%
\providecommand \doibase [0]{http://dx.doi.org/}%
\providecommand \selectlanguage [0]{\@gobble}%
\providecommand \bibinfo  [0]{\@secondoftwo}%
\providecommand \bibfield  [0]{\@secondoftwo}%
\providecommand \translation [1]{[#1]}%
\providecommand \BibitemOpen [0]{}%
\providecommand \bibitemStop [0]{}%
\providecommand \bibitemNoStop [0]{.\EOS\space}%
\providecommand \EOS [0]{\spacefactor3000\relax}%
\providecommand \BibitemShut  [1]{\csname bibitem#1\endcsname}%
\let\auto@bib@innerbib\@empty
\bibitem [{\citenamefont {Behrisch}\ and\ \citenamefont
  {Eckstein}(2007)}]{Behrisch2007}%
  \BibitemOpen
  \bibfield  {author} {\bibinfo {author} {\bibfnamefont {R.}~\bibnamefont
  {Behrisch}}\ and\ \bibinfo {author} {\bibfnamefont {W.}~\bibnamefont
  {Eckstein}},\ }\href {https://www.springer.com/gp/book/9783540445005} {\emph
  {\bibinfo {title} {{Sputtering by Particle Bombardment}}}}\ (\bibinfo
  {publisher} {Springer},\ \bibinfo {year} {2007})\BibitemShut {NoStop}%
\bibitem [{\citenamefont {Goehlich}\ and\ \citenamefont
  {D{\"{o}}bele}(1996)}]{Goehlich1996}%
  \BibitemOpen
  \bibfield  {author} {\bibinfo {author} {\bibfnamefont {A.}~\bibnamefont
  {Goehlich}}\ and\ \bibinfo {author} {\bibfnamefont {H.~F.}\ \bibnamefont
  {D{\"{o}}bele}},\ }\bibfield  {title} {\enquote {\bibinfo {title} {{Angle
  resolved velocity distributions of sputtered aluminum atoms}},}\ }\href
  {\doibase 10.1016/0168-583X(95)01448-9} {\bibfield  {journal} {\bibinfo
  {journal} {Nuclear Instruments and Methods in Physics Research Section B:
  Beam Interactions with Materials and Atoms}\ }\textbf {\bibinfo {volume}
  {115}},\ \bibinfo {pages} {489--492} (\bibinfo {year} {1996})}\BibitemShut
  {NoStop}%
\bibitem [{\citenamefont {Franke}\ \emph {et~al.}(1997)\citenamefont {Franke},
  \citenamefont {Neumann}, \citenamefont {Zeuner}, \citenamefont {Frank},\ and\
  \citenamefont {Bigl}}]{Franke1997}%
  \BibitemOpen
  \bibfield  {author} {\bibinfo {author} {\bibfnamefont {E.}~\bibnamefont
  {Franke}}, \bibinfo {author} {\bibfnamefont {H.}~\bibnamefont {Neumann}},
  \bibinfo {author} {\bibfnamefont {M.}~\bibnamefont {Zeuner}}, \bibinfo
  {author} {\bibfnamefont {W.}~\bibnamefont {Frank}}, \ and\ \bibinfo {author}
  {\bibfnamefont {F.}~\bibnamefont {Bigl}},\ }\bibfield  {title} {\enquote
  {\bibinfo {title} {{Particle energy and angle distributions in ion beam
  sputtering}},}\ }\href {\doibase 10.1016/S0257-8972(97)00304-6} {\bibfield
  {journal} {\bibinfo  {journal} {Surface and Coatings Technology}\ }\textbf
  {\bibinfo {volume} {97}},\ \bibinfo {pages} {90--96} (\bibinfo {year}
  {1997})}\BibitemShut {NoStop}%
\bibitem [{\citenamefont {Goehlich}, \citenamefont {Niem{\"{o}}ller},\ and\
  \citenamefont {D{\"{o}}bele}(1997)}]{Goehlich1997}%
  \BibitemOpen
  \bibfield  {author} {\bibinfo {author} {\bibfnamefont {A.}~\bibnamefont
  {Goehlich}}, \bibinfo {author} {\bibfnamefont {N.}~\bibnamefont
  {Niem{\"{o}}ller}}, \ and\ \bibinfo {author} {\bibfnamefont {H.~F.}\
  \bibnamefont {D{\"{o}}bele}},\ }\bibfield  {title} {\enquote {\bibinfo
  {title} {{Angle resolved velocity distributions of sputtered medium Z
  atoms}},}\ }\href {\doibase 10.1016/S0022-3115(97)80213-9} {\bibfield
  {journal} {\bibinfo  {journal} {Journal of Nuclear Materials}\ }\textbf
  {\bibinfo {volume} {241-243}},\ \bibinfo {pages} {1160--1163} (\bibinfo
  {year} {1997})}\BibitemShut {NoStop}%
\bibitem [{\citenamefont {Goehlich}, \citenamefont {Niem{\"{o}}ller},\ and\
  \citenamefont {D{\"{o}}bele}(1999)}]{Goehlich1999}%
  \BibitemOpen
  \bibfield  {author} {\bibinfo {author} {\bibfnamefont {A.}~\bibnamefont
  {Goehlich}}, \bibinfo {author} {\bibfnamefont {N.}~\bibnamefont
  {Niem{\"{o}}ller}}, \ and\ \bibinfo {author} {\bibfnamefont {H.~F.}\
  \bibnamefont {D{\"{o}}bele}},\ }\bibfield  {title} {\enquote {\bibinfo
  {title} {{Determination of angle resolved velocity distributions of sputtered
  tungsten atoms}},}\ }\href {\doibase 10.1016/S0022-3115(98)00830-7}
  {\bibfield  {journal} {\bibinfo  {journal} {Journal of Nuclear Materials}\
  }\textbf {\bibinfo {volume} {266-269}},\ \bibinfo {pages} {501--506}
  (\bibinfo {year} {1999})}\BibitemShut {NoStop}%
\bibitem [{\citenamefont {Goehlich}, \citenamefont {Gillmann},\ and\
  \citenamefont {D{\"{o}}bele}(2000)}]{Goehlich2000}%
  \BibitemOpen
  \bibfield  {author} {\bibinfo {author} {\bibfnamefont {A.}~\bibnamefont
  {Goehlich}}, \bibinfo {author} {\bibfnamefont {D.}~\bibnamefont {Gillmann}},
  \ and\ \bibinfo {author} {\bibfnamefont {H.~F.}\ \bibnamefont
  {D{\"{o}}bele}},\ }\bibfield  {title} {\enquote {\bibinfo {title} {{Angular
  resolved energy distributions of sputtered atoms at low bombarding
  energy}},}\ }\href {\doibase 10.1016/S0168-583X(99)01106-4} {\bibfield
  {journal} {\bibinfo  {journal} {Nuclear Instruments and Methods in Physics
  Research Section B: Beam Interactions with Materials and Atoms}\ }\textbf
  {\bibinfo {volume} {164-165}},\ \bibinfo {pages} {834--839} (\bibinfo {year}
  {2000})}\BibitemShut {NoStop}%
\bibitem [{\citenamefont {Goehlich}, \citenamefont {Niem{\"{o}}ller},\ and\
  \citenamefont {D{\"{o}}bele}(2000)}]{Goehlich2000a}%
  \BibitemOpen
  \bibfield  {author} {\bibinfo {author} {\bibfnamefont {A.}~\bibnamefont
  {Goehlich}}, \bibinfo {author} {\bibfnamefont {N.}~\bibnamefont
  {Niem{\"{o}}ller}}, \ and\ \bibinfo {author} {\bibfnamefont {H.~F.}\
  \bibnamefont {D{\"{o}}bele}},\ }\bibfield  {title} {\enquote {\bibinfo
  {title} {{Anisotropy effects in physical sputtering investigated by
  laser-induced fluorescence spectroscopy}},}\ }\href {\doibase
  10.1103/PhysRevB.62.9349} {\bibfield  {journal} {\bibinfo  {journal} {Phys.
  Rev. B}\ }\textbf {\bibinfo {volume} {62}},\ \bibinfo {pages} {9349--9358}
  (\bibinfo {year} {2000})}\BibitemShut {NoStop}%
\bibitem [{\citenamefont {Goehlich}, \citenamefont {Gillmann},\ and\
  \citenamefont {D{\"{o}}bele}(2001)}]{Goehlich2001}%
  \BibitemOpen
  \bibfield  {author} {\bibinfo {author} {\bibfnamefont {A.}~\bibnamefont
  {Goehlich}}, \bibinfo {author} {\bibfnamefont {D.}~\bibnamefont {Gillmann}},
  \ and\ \bibinfo {author} {\bibfnamefont {H.~F.}\ \bibnamefont
  {D{\"{o}}bele}},\ }\bibfield  {title} {\enquote {\bibinfo {title} {{An
  experimental investigation of angular resolved energy distributions of atoms
  sputtered from evaporated aluminum films}},}\ }\href {\doibase
  10.1016/S0168-583X(01)00573-0} {\bibfield  {journal} {\bibinfo  {journal}
  {Nuclear Instruments and Methods in Physics Research Section B: Beam
  Interactions with Materials and Atoms}\ }\textbf {\bibinfo {volume} {179}},\
  \bibinfo {pages} {351--363} (\bibinfo {year} {2001})}\BibitemShut {NoStop}%
\bibitem [{\citenamefont {Stepanova}\ and\ \citenamefont
  {Dew}(2001)}]{Stepanova2001}%
  \BibitemOpen
  \bibfield  {author} {\bibinfo {author} {\bibfnamefont {M.}~\bibnamefont
  {Stepanova}}\ and\ \bibinfo {author} {\bibfnamefont {S.~K.}\ \bibnamefont
  {Dew}},\ }\bibfield  {title} {\enquote {\bibinfo {title} {{Estimates of
  differential sputtering yields for deposition applications}},}\ }\href
  {\doibase 10.1116/1.1405515} {\bibfield  {journal} {\bibinfo  {journal}
  {Journal of Vacuum Science {\&} Technology A}\ }\textbf {\bibinfo {volume}
  {19}},\ \bibinfo {pages} {2805--2816} (\bibinfo {year} {2001})}\BibitemShut
  {NoStop}%
\bibitem [{\citenamefont {Stepanova}\ and\ \citenamefont
  {Dew}(2002)}]{Stepanova2002}%
  \BibitemOpen
  \bibfield  {author} {\bibinfo {author} {\bibfnamefont {M.}~\bibnamefont
  {Stepanova}}\ and\ \bibinfo {author} {\bibfnamefont {S.~K.}\ \bibnamefont
  {Dew}},\ }\bibfield  {title} {\enquote {\bibinfo {title} {{Discrete-path
  transport theory of physical sputtering}},}\ }\href {\doibase
  10.1063/1.1488245} {\bibfield  {journal} {\bibinfo  {journal} {Journal of
  Applied Physics}\ }\textbf {\bibinfo {volume} {92}},\ \bibinfo {pages}
  {1699--1708} (\bibinfo {year} {2002})}\BibitemShut {NoStop}%
\bibitem [{\citenamefont {Stepanova}, \citenamefont {Dew},\ and\ \citenamefont
  {Soshnikov}(2002)}]{Stepanova2002a}%
  \BibitemOpen
  \bibfield  {author} {\bibinfo {author} {\bibfnamefont {M.}~\bibnamefont
  {Stepanova}}, \bibinfo {author} {\bibfnamefont {S.~K.}\ \bibnamefont {Dew}},
  \ and\ \bibinfo {author} {\bibfnamefont {I.~P.}\ \bibnamefont {Soshnikov}},\
  }\bibfield  {title} {\enquote {\bibinfo {title} {{Sputtering from
  ion-beam-roughened Cu surfaces}},}\ }\href {\doibase
  10.1103/PhysRevB.66.125407} {\bibfield  {journal} {\bibinfo  {journal} {Phys.
  Rev. B}\ }\textbf {\bibinfo {volume} {66}},\ \bibinfo {pages} {125407}
  (\bibinfo {year} {2002})}\BibitemShut {NoStop}%
\bibitem [{\citenamefont {Stepanova}\ and\ \citenamefont
  {Dew}(2004)}]{Stepanova2004}%
  \BibitemOpen
  \bibfield  {author} {\bibinfo {author} {\bibfnamefont {M.}~\bibnamefont
  {Stepanova}}\ and\ \bibinfo {author} {\bibfnamefont {S.~K.}\ \bibnamefont
  {Dew}},\ }\bibfield  {title} {\enquote {\bibinfo {title} {{Anisotropic
  energies of sputtered atoms under oblique ion incidence}},}\ }\href {\doibase
  10.1016/j.nimb.2003.09.013} {\bibfield  {journal} {\bibinfo  {journal}
  {Nuclear Instruments and Methods in Physics Research Section B: Beam
  Interactions with Materials and Atoms}\ }\textbf {\bibinfo {volume} {215}},\
  \bibinfo {pages} {357--365} (\bibinfo {year} {2004})}\BibitemShut {NoStop}%
\bibitem [{\citenamefont {Thompson}(1968)}]{Thompson1968}%
  \BibitemOpen
  \bibfield  {author} {\bibinfo {author} {\bibfnamefont {M.~W.}\ \bibnamefont
  {Thompson}},\ }\bibfield  {title} {\enquote {\bibinfo {title} {{II. The
  energy spectrum of ejected atoms during the high energy sputtering of
  gold}},}\ }\href {\doibase 10.1080/14786436808227358} {\bibfield  {journal}
  {\bibinfo  {journal} {The Philosophical Magazine: A Journal of Theoretical
  Experimental and Applied Physics}\ }\textbf {\bibinfo {volume} {18}},\
  \bibinfo {pages} {377--414} (\bibinfo {year} {1968})}\BibitemShut {NoStop}%
\bibitem [{\citenamefont {Sigmund}(1968)}]{Sigmund1969}%
  \BibitemOpen
  \bibfield  {author} {\bibinfo {author} {\bibfnamefont {P.}~\bibnamefont
  {Sigmund}},\ }\bibfield  {title} {\enquote {\bibinfo {title} {{Theory of
  Sputtering. I. Sputtering Yield of Amorphous and Polycrystalline Targets}},}\
  }\href {\doibase 10.1103/PhysRev.184.383} {\bibfield  {journal} {\bibinfo
  {journal} {Physical Review}\ }\textbf {\bibinfo {volume} {184}},\ \bibinfo
  {pages} {383} (\bibinfo {year} {1968})}\BibitemShut {NoStop}%
\bibitem [{\citenamefont {Lautenschl{\"{a}}ger}\ \emph
  {et~al.}(2017)\citenamefont {Lautenschl{\"{a}}ger}, \citenamefont {Feder},
  \citenamefont {Neumann}, \citenamefont {Rice}, \citenamefont {Schubert},\
  and\ \citenamefont {Bundesmann}}]{Lautenschlager2017}%
  \BibitemOpen
  \bibfield  {author} {\bibinfo {author} {\bibfnamefont {T.}~\bibnamefont
  {Lautenschl{\"{a}}ger}}, \bibinfo {author} {\bibfnamefont {R.}~\bibnamefont
  {Feder}}, \bibinfo {author} {\bibfnamefont {H.}~\bibnamefont {Neumann}},
  \bibinfo {author} {\bibfnamefont {C.}~\bibnamefont {Rice}}, \bibinfo {author}
  {\bibfnamefont {M.}~\bibnamefont {Schubert}}, \ and\ \bibinfo {author}
  {\bibfnamefont {C.}~\bibnamefont {Bundesmann}},\ }\bibfield  {title}
  {\enquote {\bibinfo {title} {{Reactive ion beam sputtering of Ti: Influence
  of process parameters on angular and energy distribution of sputtered and
  backscattered particles}},}\ }\href {\doibase 10.1016/j.nimb.2016.08.017}
  {\bibfield  {journal} {\bibinfo  {journal} {Journal of Vacuum Science {\&}
  Technology A: Vacuum, Surfaces, and Films}\ }\textbf {\bibinfo {volume}
  {35}},\ \bibinfo {pages} {041001} (\bibinfo {year} {2017})}\BibitemShut
  {NoStop}%
\bibitem [{\citenamefont {Mateev}\ \emph {et~al.}(2018)\citenamefont {Mateev},
  \citenamefont {Lautenschl{\"{a}}ger}, \citenamefont {Spemann}, \citenamefont
  {Finzel}, \citenamefont {Gerlach}, \citenamefont {Frost},\ and\ \citenamefont
  {Bundesmann}}]{Mateev2018}%
  \BibitemOpen
  \bibfield  {author} {\bibinfo {author} {\bibfnamefont {M.}~\bibnamefont
  {Mateev}}, \bibinfo {author} {\bibfnamefont {T.}~\bibnamefont
  {Lautenschl{\"{a}}ger}}, \bibinfo {author} {\bibfnamefont {D.}~\bibnamefont
  {Spemann}}, \bibinfo {author} {\bibfnamefont {A.}~\bibnamefont {Finzel}},
  \bibinfo {author} {\bibfnamefont {J.~W.}\ \bibnamefont {Gerlach}}, \bibinfo
  {author} {\bibfnamefont {F.}~\bibnamefont {Frost}}, \ and\ \bibinfo {author}
  {\bibfnamefont {C.}~\bibnamefont {Bundesmann}},\ }\bibfield  {title}
  {\enquote {\bibinfo {title} {{Systematic investigation of the reactive ion
  beam sputter deposition process of SiO2}},}\ }\href {\doibase
  10.1140/epjb/e2018-80453-x} {\bibfield  {journal} {\bibinfo  {journal}
  {European Physical Journal B}\ }\textbf {\bibinfo {volume} {91}},\ \bibinfo
  {pages} {45} (\bibinfo {year} {2018})}\BibitemShut {NoStop}%
\bibitem [{\citenamefont {Bundesmann}\ and\ \citenamefont
  {Neumann}(2018)}]{Bundesmann2018}%
  \BibitemOpen
  \bibfield  {author} {\bibinfo {author} {\bibfnamefont {C.}~\bibnamefont
  {Bundesmann}}\ and\ \bibinfo {author} {\bibfnamefont {H.}~\bibnamefont
  {Neumann}},\ }\bibfield  {title} {\enquote {\bibinfo {title} {{Tutorial: The
  systematics of ion beam sputtering for deposition of thin films with tailored
  properties}},}\ }\href {\doibase 10.1063/1.5054046} {\bibfield  {journal}
  {\bibinfo  {journal} {Journal of Applied Physics}\ }\textbf {\bibinfo
  {volume} {124}},\ \bibinfo {pages} {231102} (\bibinfo {year}
  {2018})}\BibitemShut {NoStop}%
\bibitem [{\citenamefont {Okutani}\ \emph {et~al.}(1980)\citenamefont
  {Okutani}, \citenamefont {Shikata}, \citenamefont {Ichimura},\ and\
  \citenamefont {Shimizu}}]{Okutani1980}%
  \BibitemOpen
  \bibfield  {author} {\bibinfo {author} {\bibfnamefont {T.}~\bibnamefont
  {Okutani}}, \bibinfo {author} {\bibfnamefont {M.}~\bibnamefont {Shikata}},
  \bibinfo {author} {\bibfnamefont {S.}~\bibnamefont {Ichimura}}, \ and\
  \bibinfo {author} {\bibfnamefont {R.}~\bibnamefont {Shimizu}},\ }\bibfield
  {title} {\enquote {\bibinfo {title} {{Angular distribution of Si atoms
  sputtered by keV Ar+ ions}},}\ }\href {\doibase 10.1063/1.327957} {\bibfield
  {journal} {\bibinfo  {journal} {Journal of Applied Physics}\ }\textbf
  {\bibinfo {volume} {51}},\ \bibinfo {pages} {2884--2887} (\bibinfo {year}
  {1980})}\BibitemShut {NoStop}%
\bibitem [{\citenamefont {Chernysh}\ \emph {et~al.}(2004)\citenamefont
  {Chernysh}, \citenamefont {Kulikauskas}, \citenamefont {Patrakeev},
  \citenamefont {Abdul-cader},\ and\ \citenamefont {Shulga}}]{Chernysh2004}%
  \BibitemOpen
  \bibfield  {author} {\bibinfo {author} {\bibfnamefont {V.~S.}\ \bibnamefont
  {Chernysh}}, \bibinfo {author} {\bibfnamefont {V.~S.}\ \bibnamefont
  {Kulikauskas}}, \bibinfo {author} {\bibfnamefont {A.~S.}\ \bibnamefont
  {Patrakeev}}, \bibinfo {author} {\bibfnamefont {K.~M.}\ \bibnamefont
  {Abdul-cader}}, \ and\ \bibinfo {author} {\bibfnamefont {V.~I.}\ \bibnamefont
  {Shulga}},\ }\bibfield  {title} {\enquote {\bibinfo {title} {{Angular
  distribution of atoms sputtered from silicon by 1–10 keV Ar ions}},}\
  }\href {\doibase 10.1080/10420150410001669613} {\bibfield  {journal}
  {\bibinfo  {journal} {Radiation Effects and Defects in Solids}\ }\textbf
  {\bibinfo {volume} {159}},\ \bibinfo {pages} {149--155} (\bibinfo {year}
  {2004})}\BibitemShut {NoStop}%
\bibitem [{\citenamefont {Wittmaack}(1984)}]{Wittmaack1984}%
  \BibitemOpen
  \bibfield  {author} {\bibinfo {author} {\bibfnamefont {K.}~\bibnamefont
  {Wittmaack}},\ }\bibfield  {title} {\enquote {\bibinfo {title} {{Angular
  dependence of secondary ion emission from silicon bombarded with inert gas
  ions}},}\ }\href {\doibase 10.1016/0168-583X(84)90290-8} {\bibfield
  {journal} {\bibinfo  {journal} {Nuclear Instruments and Methods in Physics
  Research Section B: Beam Interactions with Materials and Atoms}\ }\textbf
  {\bibinfo {volume} {2}},\ \bibinfo {pages} {674--678} (\bibinfo {year}
  {1984})}\BibitemShut {NoStop}%
\bibitem [{\citenamefont {Mutzke}\ \emph {et~al.}(2011)\citenamefont {Mutzke},
  \citenamefont {Schneider}, \citenamefont {Eckstein},\ and\ \citenamefont
  {Dohmen}}]{Mutzke2011}%
  \BibitemOpen
  \bibfield  {author} {\bibinfo {author} {\bibfnamefont {A.}~\bibnamefont
  {Mutzke}}, \bibinfo {author} {\bibfnamefont {R.}~\bibnamefont {Schneider}},
  \bibinfo {author} {\bibfnamefont {W.}~\bibnamefont {Eckstein}}, \ and\
  \bibinfo {author} {\bibfnamefont {R.}~\bibnamefont {Dohmen}},\ }\href
  {http://hdl.handle.net/11858/00-001M-0000-0026-EAF9-A} {\enquote {\bibinfo
  {title} {{SDTrimSP Version 5.00, IPP report 12/8}},}\ }\bibinfo {type} {Tech.
  Rep.}\ (\bibinfo  {institution} {Max-Planck-Institut f{\"{u}}r Plasmaphysik,
  Garching},\ \bibinfo {year} {2011})\BibitemShut {NoStop}%
\bibitem [{\citenamefont {Zeuner}\ \emph {et~al.}(2001)\citenamefont {Zeuner},
  \citenamefont {Scholze}, \citenamefont {Dathe},\ and\ \citenamefont
  {Neumann}}]{Zeuner2001}%
  \BibitemOpen
  \bibfield  {author} {\bibinfo {author} {\bibfnamefont {M.}~\bibnamefont
  {Zeuner}}, \bibinfo {author} {\bibfnamefont {F.}~\bibnamefont {Scholze}},
  \bibinfo {author} {\bibfnamefont {B.}~\bibnamefont {Dathe}}, \ and\ \bibinfo
  {author} {\bibfnamefont {H.}~\bibnamefont {Neumann}},\ }\bibfield  {title}
  {\enquote {\bibinfo {title} {{Optimisation and characterisation of a TCP type
  RF broad beam ion source}},}\ }\href {\doibase 10.1016/S0257-8972(01)01219-1}
  {\bibfield  {journal} {\bibinfo  {journal} {Surface and Coatings Technology}\
  }\textbf {\bibinfo {volume} {142-144}},\ \bibinfo {pages} {39--48} (\bibinfo
  {year} {2001})}\BibitemShut {NoStop}%
\bibitem [{\citenamefont {Wilson}, \citenamefont {Haggmark},\ and\
  \citenamefont {Biersack}(1977)}]{Wilson1977}%
  \BibitemOpen
  \bibfield  {author} {\bibinfo {author} {\bibfnamefont {W.~D.}\ \bibnamefont
  {Wilson}}, \bibinfo {author} {\bibfnamefont {L.~G.}\ \bibnamefont
  {Haggmark}}, \ and\ \bibinfo {author} {\bibfnamefont {J.~P.}\ \bibnamefont
  {Biersack}},\ }\bibfield  {title} {\enquote {\bibinfo {title} {{Calculations
  of nuclear stopping, ranges, and straggling in the low-energy region}},}\
  }\href {\doibase 10.1103/PhysRevB.15.2458} {\bibfield  {journal} {\bibinfo
  {journal} {Phys. Rev. B}\ }\textbf {\bibinfo {volume} {15}},\ \bibinfo
  {pages} {2458--2468} (\bibinfo {year} {1977})}\BibitemShut {NoStop}%
\bibitem [{\citenamefont {Lindhard}\ and\ \citenamefont
  {Scharff}(1961)}]{Lindhard1961}%
  \BibitemOpen
  \bibfield  {author} {\bibinfo {author} {\bibfnamefont {J.}~\bibnamefont
  {Lindhard}}\ and\ \bibinfo {author} {\bibfnamefont {M.}~\bibnamefont
  {Scharff}},\ }\bibfield  {title} {\enquote {\bibinfo {title} {{Energy
  Dissipation by Ions in the kev Region}},}\ }\href {\doibase
  10.1103/PhysRev.124.128} {\bibfield  {journal} {\bibinfo  {journal} {Phys.
  Rev.}\ }\textbf {\bibinfo {volume} {124}},\ \bibinfo {pages} {128--130}
  (\bibinfo {year} {1961})}\BibitemShut {NoStop}%
\bibitem [{\citenamefont {Oen}\ and\ \citenamefont {Robinson}(1976)}]{Oen1976}%
  \BibitemOpen
  \bibfield  {author} {\bibinfo {author} {\bibfnamefont {O.~S.}\ \bibnamefont
  {Oen}}\ and\ \bibinfo {author} {\bibfnamefont {M.~T.}\ \bibnamefont
  {Robinson}},\ }\bibfield  {title} {\enquote {\bibinfo {title} {{Computer
  studies of the reflection of light ions from solids}},}\ }\href {\doibase
  10.1016/0029-554X(76)90806-5} {\bibfield  {journal} {\bibinfo  {journal}
  {Nuclear Instruments and Methods}\ }\textbf {\bibinfo {volume} {132}},\
  \bibinfo {pages} {647--653} (\bibinfo {year} {1976})}\BibitemShut {NoStop}%
\bibitem [{\citenamefont {Feder}\ \emph {et~al.}(2013)\citenamefont {Feder},
  \citenamefont {Bundesmann}, \citenamefont {Neumann},\ and\ \citenamefont
  {Rauschenbach}}]{Feder2013}%
  \BibitemOpen
  \bibfield  {author} {\bibinfo {author} {\bibfnamefont {R.}~\bibnamefont
  {Feder}}, \bibinfo {author} {\bibfnamefont {C.}~\bibnamefont {Bundesmann}},
  \bibinfo {author} {\bibfnamefont {H.}~\bibnamefont {Neumann}}, \ and\
  \bibinfo {author} {\bibfnamefont {B.}~\bibnamefont {Rauschenbach}},\
  }\bibfield  {title} {\enquote {\bibinfo {title} {{Ion beam sputtering of Ag -
  Angular and energetic distributions of sputtered and scattered particles}},}\
  }\href {\doibase 10.1016/j.nimb.2013.09.007} {\bibfield  {journal} {\bibinfo
  {journal} {Nuclear Instruments and Methods in Physics Research, Section B:
  Beam Interactions with Materials and Atoms}\ }\textbf {\bibinfo {volume}
  {316}},\ \bibinfo {pages} {198--204} (\bibinfo {year} {2013})}\BibitemShut
  {NoStop}%
\bibitem [{\citenamefont {Bundesmann}\ \emph {et~al.}(2015)\citenamefont
  {Bundesmann}, \citenamefont {Feder}, \citenamefont {Lautenschl{\"{a}}ger},\
  and\ \citenamefont {Neumann}}]{Bundesmann2015}%
  \BibitemOpen
  \bibfield  {author} {\bibinfo {author} {\bibfnamefont {C.}~\bibnamefont
  {Bundesmann}}, \bibinfo {author} {\bibfnamefont {R.}~\bibnamefont {Feder}},
  \bibinfo {author} {\bibfnamefont {T.}~\bibnamefont {Lautenschl{\"{a}}ger}}, \
  and\ \bibinfo {author} {\bibfnamefont {H.}~\bibnamefont {Neumann}},\
  }\bibfield  {title} {\enquote {\bibinfo {title} {{Energy Distribution of
  Secondary Particles in Ion Beam Deposition Process of Ag: Experiment,
  Calculation and Simulation}},}\ }\href {\doibase 10.1002/ctpp.201510015}
  {\bibfield  {journal} {\bibinfo  {journal} {Contributions to Plasma Physics}\
  }\textbf {\bibinfo {volume} {55}},\ \bibinfo {pages} {737--746} (\bibinfo
  {year} {2015})}\BibitemShut {NoStop}%
\bibitem [{\citenamefont {Feder}\ \emph {et~al.}(2014)\citenamefont {Feder},
  \citenamefont {Bundesmann}, \citenamefont {Neumann},\ and\ \citenamefont
  {Rauschenbach}}]{Feder2014}%
  \BibitemOpen
  \bibfield  {author} {\bibinfo {author} {\bibfnamefont {R.}~\bibnamefont
  {Feder}}, \bibinfo {author} {\bibfnamefont {C.}~\bibnamefont {Bundesmann}},
  \bibinfo {author} {\bibfnamefont {H.}~\bibnamefont {Neumann}}, \ and\
  \bibinfo {author} {\bibfnamefont {B.}~\bibnamefont {Rauschenbach}},\
  }\bibfield  {title} {\enquote {\bibinfo {title} {{Ion beam sputtering of
  germanium - Energy and angular distribution of sputtered and scattered
  particles}},}\ }\href {\doibase 10.1016/j.nimb.2014.05.009} {\bibfield
  {journal} {\bibinfo  {journal} {Nuclear Instruments and Methods in Physics
  Research, Section B: Beam Interactions with Materials and Atoms}\ }\textbf
  {\bibinfo {volume} {334}},\ \bibinfo {pages} {88--95} (\bibinfo {year}
  {2014})}\BibitemShut {NoStop}%
\bibitem [{\citenamefont {Lautenschl{\"{a}}ger}\ \emph
  {et~al.}(2016)\citenamefont {Lautenschl{\"{a}}ger}, \citenamefont {Feder},
  \citenamefont {Neumann}, \citenamefont {Rice}, \citenamefont {Schubert},\
  and\ \citenamefont {Bundesmann}}]{Lautenschlager2016}%
  \BibitemOpen
  \bibfield  {author} {\bibinfo {author} {\bibfnamefont {T.}~\bibnamefont
  {Lautenschl{\"{a}}ger}}, \bibinfo {author} {\bibfnamefont {R.}~\bibnamefont
  {Feder}}, \bibinfo {author} {\bibfnamefont {H.}~\bibnamefont {Neumann}},
  \bibinfo {author} {\bibfnamefont {C.}~\bibnamefont {Rice}}, \bibinfo {author}
  {\bibfnamefont {M.}~\bibnamefont {Schubert}}, \ and\ \bibinfo {author}
  {\bibfnamefont {C.}~\bibnamefont {Bundesmann}},\ }\bibfield  {title}
  {\enquote {\bibinfo {title} {{Ion beam sputtering of Ti: Influence of process
  parameters on angular and energy distribution of sputtered and backscattered
  particles}},}\ }\href {\doibase 10.1016/j.nimb.2016.08.017} {\bibfield
  {journal} {\bibinfo  {journal} {Nuclear Instruments and Methods in Physics
  Research B}\ }\textbf {\bibinfo {volume} {385}},\ \bibinfo {pages} {30--39}
  (\bibinfo {year} {2016})}\BibitemShut {NoStop}%
\bibitem [{\citenamefont {Ziegler}, \citenamefont {Biersack},\ and\
  \citenamefont {Ziegler}(2013)}]{SRIM2013}%
  \BibitemOpen
  \bibfield  {author} {\bibinfo {author} {\bibfnamefont {J.~F.}\ \bibnamefont
  {Ziegler}}, \bibinfo {author} {\bibfnamefont {J.~P.}\ \bibnamefont
  {Biersack}}, \ and\ \bibinfo {author} {\bibfnamefont {M.~D.}\ \bibnamefont
  {Ziegler}},\ }\bibfield  {title} {\enquote {\bibinfo {title} {{SRIM - The
  Stopping and Range of Ions in Matter}},}\ }\href {http://www.srim.org/}
  {\bibfield  {journal} {\bibinfo  {journal} {http://www.srim.org/}\ }
  (\bibinfo {year} {2013})}\BibitemShut {NoStop}%
\bibitem [{\citenamefont {Roosendaal}\ and\ \citenamefont
  {Sanders}(1980)}]{Roosendaal1980}%
  \BibitemOpen
  \bibfield  {author} {\bibinfo {author} {\bibfnamefont {H.~E.}\ \bibnamefont
  {Roosendaal}}\ and\ \bibinfo {author} {\bibfnamefont {J.~B.}\ \bibnamefont
  {Sanders}},\ }\bibfield  {title} {\enquote {\bibinfo {title} {{On the energy
  distribution and angular distribution of sputtered particles}},}\ }\href
  {\doibase 10.1080/00337578008210025} {\bibfield  {journal} {\bibinfo
  {journal} {Radiation Effects}\ }\textbf {\bibinfo {volume} {52}},\ \bibinfo
  {pages} {137--143} (\bibinfo {year} {1980})}\BibitemShut {NoStop}%
\end{thebibliography}%

\end{document}